\documentclass[
reprint,
superscriptaddress,
nofootinbib,
amsmath,amssymb,
prl,
]{revtex4-1}

\usepackage[utf8]{inputenc}
\usepackage{amsmath}
\usepackage{mathtools}
\usepackage{amssymb,amstext,amsfonts}
\usepackage{hyperref}
\usepackage{graphicx}
\usepackage{float}
\usepackage{microtype}
\usepackage{subfigure}
\usepackage{multirow}
\usepackage{lineno}
\usepackage{color}
\usepackage{comment}
\usepackage[normalem]{ulem}

\newcommand{\be}{ \begin{equation} }
\newcommand{\ee}{ \end{equation}}

\renewcommand{\vec}[1]{\ensuremath{\mathbf{#1}}}

\begin{document}

\title{Observational Black Hole Spectroscopy:\\A time-domain multimode analysis of GW150914}

\author{Gregorio Carullo}
\affiliation{Dipartimento di Fisica ``Enrico Fermi'', Universit\`a di Pisa, Pisa I-56127, Italy}
\affiliation{INFN sezione di Pisa, Pisa I-56127, Italy}
\author{Walter Del Pozzo}
\affiliation{Dipartimento di Fisica ``Enrico Fermi'', Universit\`a di Pisa, Pisa I-56127, Italy}
\affiliation{INFN sezione di Pisa, Pisa I-56127, Italy}
\author{John Veitch}
\affiliation{Institute for Gravitational Research, University of Glasgow, Glasgow, G12 8QQ, United Kingdom}

\date{\today}

\begin{abstract}
The detection of the least damped quasi-normal mode from the remnant of the gravitational
wave event GW150914 realised the long sought possibility to observationally study the properties of 
quasi-stationary black hole spacetimes through gravitational waves. Past literature has extensively 
explored this possibility and the emerging field has been 
named ``black hole spectroscopy''.
In this study, we present results regarding the ringdown spectrum of GW150914,
obtained by application of Bayesian inference to identify and characterise the ringdown modes.
We employ a pure time-domain analysis method which infers from the data the time of transition between 
the non-linear and quasi-linear regime of the post-merger emission 
in concert with all other parameters characterising the source. 
We find that the data provides no evidence for the presence of more than one quasi-normal mode. 
However, from the central frequency and damping time posteriors alone, no unambiguous identification of a single mode is possible.
More in-depth analysis adopting a ringdown model based on results in perturbation theory over the Kerr
metric, confirms that the data do not provide enough evidence to discriminate among an $l=2$ and the 
$l=3$ subset of modes.
Our work provides the first comprehensive agnostic framework to observationally investigate astrophysical 
black holes' ringdown spectra.

\end{abstract}

\maketitle


\section{Introduction} 

GW150914~\cite{GW150914}, the loudest binary black hole (BBH) detected so far, 
provided evidence for the presence of a ringdown at the end of the coalescence. 
The ringdown spectrum, a superposition of quasi-normal modes (QNMs)
and late-time power-law tails~\cite{Vishveshwara, Press:1971wr, Chandrasekhar:1975zza, PhysRevD.5.2439}, 
directly ties to fundamental properties of the underlying spacetime. 
Due to the final state conjecture (i.e. no-hair theorems plus the conjecture that the Kerr solution is a dynamical attractor for BH spacetimes in astrophysical scenarios) \cite{Ginzburg,Zeldovich,Israel,Carter,Hawking1972,Robinson,Mazur,Bunting, Pretorius2017, Dafermos:2008en} 
the physical spectrum of QNMs is exclusively determined by the asymptotic black hole (BH) mass and spin, 
hence ringdown observations of astrophysical BHs have the potential of verifying the Kerr nature of these objects.
Moreover, an accurate determination of the BH ringdown spectrum represents one of the most promising 
avenues to unveil BH horizon quantum effects, discovering exotic compact objects, hairy BHs or even wormholes~\cite{Cardoso_BH_mimick, Hod:2011aa, Raposo:2018xkf}. 
Several authors proposed methods and offered predictions on the feasibility of what has come 
to be known as ``black hole spectroscopy''~\cite{Dreyer:2003bv, LISA_spectroscopy, Brito:2018rfr}. 
The improvements gained by coherently combining information from multiple events have also been investigated, 
see e.g. Ref.~\cite{Yang:2017zxs}.

The first determination of a compact object QNM~\cite{TGR_paper_GW150914} by the LIGO and Virgo Collaborations (LVC)~\citep{Aasi:2013wya, TheLIGOScientific:2014jea, TheVirgo:2014hva} revealed the intrinsic difficulties in determining the transition between the non-linear merger regime to the quasi-linear one, 
where the results of BH perturbation theory are applicable.
Several studies based on numerical relativity simulations~\cite{Swetha, Carullo:2018sfu},
have proposed a start time of $\sim 15$\,M -- M being the remnant mass -- after the peak 
strain of the waveform (for an alternative approach see Ref.~\cite{Area_test}).
However, to validate these claims and to perform theory-agnostic measurements of the BH spectrum, 
we require the ability to measure the start time from the data.
With this, we can robustly test the remarkable predictions of GR, such
as the black hole area increase law \cite{Area_test}. 

In this paper, we present the first comprehensive spectroscopic analysis of the GW150914 ringdown signal that,
by operating directly in time domain, successfully identifies from the data the time of transition as well as
the most probable subset of QNMs in the data. This is achieved employing a generic 
damped sinusoids ringdown model which does not include GR predictions for its complex frequencies, 
hence generic enough to incorporate the emission of alternative compact objects possibly mimicking ringdown signals. 
We find no evidence in support of the presence of a second 
mode in addition to the one already identified in the LVC analysis. The central frequency and
damping time measured with the generic model indicate that the most-probable modes can
be identified with a subset of the the $\ell = 2$ and $\ell = 3$ modes as predicted 
by a Kerr solution~\cite{Kerr:1963ud}. 
We refine our results by adopting a model based 
on the theoretical spectrum of a Kerr BH. We demonstrate, in agreement with the generic approach, that 
the $(\ell,m,n) = (3,-3,0), (3,-2,0), (2,1,0)$ and $(2,2,0)$ modes are consistent with the remnant mass and spin of GW150914. 
The inferred transition time is consistent with the result obtained with the previous model.
Unless explicitly noted, all statistical bounds reported are $90\%$ credible regions.


\section{Time domain analysis} 
We perform our analysis in the time domain and use data from the Advanced LIGO detectors, provided by the
Gravitational-Wave Open Science Center\cite{GWOSC,Vallisneri:2014vxa}.
We model the detector noise as a wide-sense stationary Gaussian process. 
We verify the validity of this assumption, confirming its validity, 
in agreement with the results published by the LVC collaboration at the time of the discovery, 
see e.g. Ref.~\cite{TGR_paper_GW150914}.  
The stochastic process describing the detector noise is thus fully described 
by its two-point autocovariance function $C(\tau)$:
\be
C(\tau) = \int{dt \, n(t) \, n(t+\tau)}\,,
\ee
which we estimate from 4096\,s of data surrounding the event. 
The 4096\,s of data sampled at a rate of 4096\, Hz are band-passed with a 4th order Butterworth filter 
in the band [20,2028] Hz, then split into $\mathcal{X}$\,second long chunks. The autocovariance is computed as the mean 
of the individual autocovariances estimated on each chunk, excluding the one containing the time of the trigger.
To validate our noise estimation method, we exploit the Wiener-Khinchin theorem and compare the 
Discrete Fourier Transform (DFT) of the autocovariance with the power spectral density (PSD) computed with the standard Welch method.
We find good agreement between the two estimates by choosing $\mathcal{X}$ = 2. Any 
choice of $\mathcal{X} > 2\,$ does not affect our conclusions.
The log-likelihood function for the observed strain series $d(t)$, given the presence of 
a GW signal $h(t)$ is:
\begin{widetext}
\begin{align*}
\log p(d|\vec{\theta} ,I) = -\frac{1}{2}
    \int{\int{dt\,d\tau \, (d(t)-h(t;\vec{\theta})) \, C^{-1}(\tau) \,
        (d(t+\tau)-h(t+\tau ; \vec{\theta}))}}, 
\end{align*}
\end{widetext}
where the domain of integration extends over the considered segment of data.
The time-domain likelihood solves several technical issues of the analysis, 
providing a convenient framework to avoid Gibbs phenomena arising from the DFT of a fast-rising template 
which can pollute the analysis of the BH spectrum.
The analysis was performed using a nested sampling algorithm~\cite{CPNest}.


\section{Agnostic analysis} 
We first perform an \textit{agnostic} analysis of the GW150914 ringdown signal,
without assuming GR predictions on the spectrum and intensity of the emission. 
Hence we relax as many assumptions as possible and
assume a model defined by a superposition of damped sinusoids:
\be\label{eq:ag_model}
h_+ - i h_{\times} = \sum_{n} \mathcal{A}_{n} \, e^{i\tilde{\omega}_{n}(t-t_n)+\phi_{n}}\,.
\ee
where $\tilde{\omega}_{n} \equiv \omega_{n} + i/\tau_n$ is the complex ringdown frequency. The parameters
$\{  \omega_n, \tau_n, \phi_n,  \mathcal{A}_{n}, t_n \}_{n\in \mathcal{N}}$ are estimated directly from the data. 
The index $n$ labels the $\mathcal{N}$ modes considered. Note that, in addition to the frequencies $\omega_{n}$ and damping times $\tau_n$, 
we also infer the amplitudes $\mathcal{A}_n$ and start times $t_n$ from the data.

The prior distribution on the intrinsic parameters was chosen to be 
uniform within the ranges: $f_n\in [100,500]$\,Hz, $\tau_n\in [0.5, 20] $\,ms,
$\log_{10}{A_n} \in[-23,-19]$, $\varphi_n \in [0,2\pi]$\,rad,
$t_n\in[3.3,6.6]$\,ms after the peak of the waveform.
The prior on the start time corresponds to the specific choice of $[10,20]\,M_f$ after the peak time
of the strain (during the analysis $t$ = 1126259462.423\,s at LIGO Hanford site was chosen, 
in agreement with Ref.~\cite{TGR_paper_GW150914}).
This choice is guided by numerical relativity studies that looked at the beginning of the 
linearized ringdown regime validity \cite{MMRDNS_paper}.
Different choices on the prior on the start time are presented in the \textit{Discussion} section.
$M_f = 68\,M_\odot$ (in geometric units) is the median value  of the estimate presented in \cite{Abbott:2016izl}.
In addition to the intrinsic model parameters, we sample
the sky position angles and polarisation, to obtain the detector strain:
\be
h(t) = F_+(\alpha, \delta, \psi) \, h_+ + F_{\times}(\alpha, \delta, \psi) \, h_{\times}
\ee
where $F_+(\alpha, \delta, \psi), F_{\times}(\alpha, \delta, \psi)$ are the detector angular response functions~\cite{AndersonCreighton}.
For all the models employed in this paper, we chose our prior distribution to be 
isotropic for the source's sky location, and uniform in the polarisation angle $\psi\in [0,\pi]$.

\begin{figure}[t]
\includegraphics[width=0.45\textwidth]{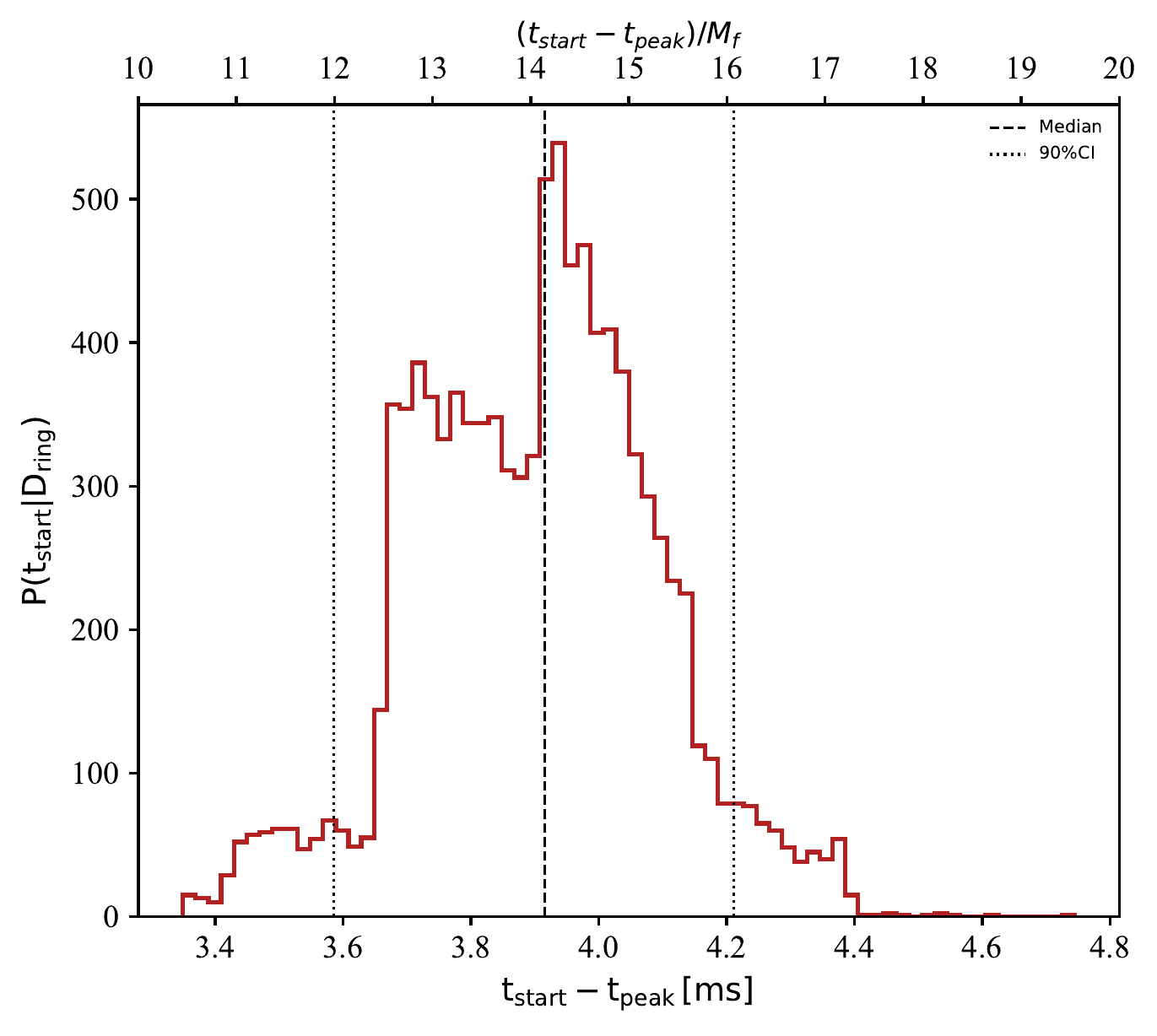}
\caption{\label{fig:time-post} Posterior distribution over $t_{\rm start}$ obtained assuming a uniform prior
distribution in $[10,20]\,M_f$.}
\end{figure}

\begin{figure}[t]
\includegraphics[width=0.45\textwidth]{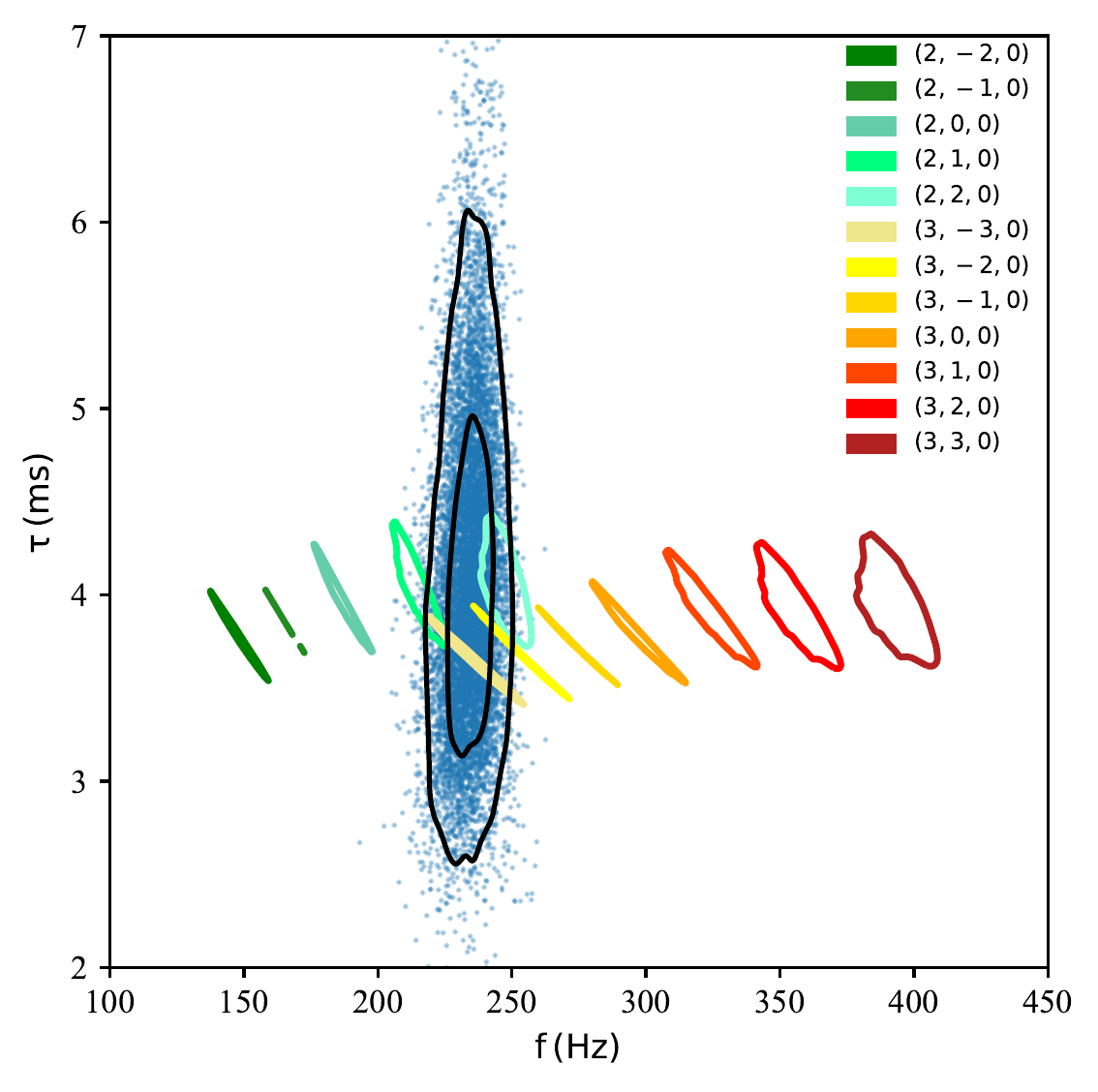}
\caption{\label{fig:Spectroscopy_plot} BH spectroscopy from the two-
dimensional posterior for central frequency and damping time obtained with a 
single damped sinusoid ringdown model. The colored contours are the  
$90 \%$ credible intervals for particular $(\ell,m,n)$ Kerr modes, derived from
the LVC reported remnant mass $M_f$ and spin $a_f$ (derived from inspiral).
We show the $\ell=2$ and $\ell=3$ modes as other $\ell$'s do not overlap
with the posterior distribution.}
\end{figure}

We begin by assuming a single mode damped sinusoid model to detect the most excited mode directly from the data.
Fig.~\ref{fig:Spectroscopy_plot} shows the joint posterior distribution for the central 
frequency $f$ and damping time $\tau$ from our analysis. We find $f=234_{-12}^{+11}\,\mathrm{Hz}$ 
and $\tau = 4.0_{-1.0}^{+1.5}\, \mathrm{ms}$, which is consistent both with the 
predicted values from a full Inspiral-Merger-Ringdown (IMR) analysis and with the 
late time ($t_{\rm start} \geq t_{\rm merger}+3$\,ms) unmodelled analyses in Refs.~\cite{TGR_paper_GW150914, Brito:2018rfr}.
To investigate which mode has the highest probability of matching the recovered unmodelled posterior \emph{a posteriori}, 
according to GR predictions, we use the samples for the progenitors masses and spins released by the LVC~\cite{O1_samples_release}, 
combined with fitting formulae obtained from numerical relativity simulations \cite{UIB2016_FF_paper, LISA_spectroscopy, Berti_website} 
and employ them to \emph{predict} the corresponding frequency and damping time for a set of modes which overlap with the unmodelled posterior.
Figure~\ref{fig:Spectroscopy_plot} shows the $90\%$ CI on $n=0, \, l=2,3$ modes 
obtained with the described procedure.
The largest overlap, quantified through Bayes' theorem, is obtained for the $\{ (2,2,0), \, (3,-3,0) \}$ modes.
We also find $\mathcal{A} = (3.22_{-1.1}^{+1.4})\times 10^{-21}$, and, most notably, we determine, directly from 
the data, $t_{\rm start} = 3.9_{-0.3}^{+0.3}$ ms, Fig.~\ref{fig:time-post}. 

\begin{figure}[h]
\includegraphics[width=0.50\textwidth]{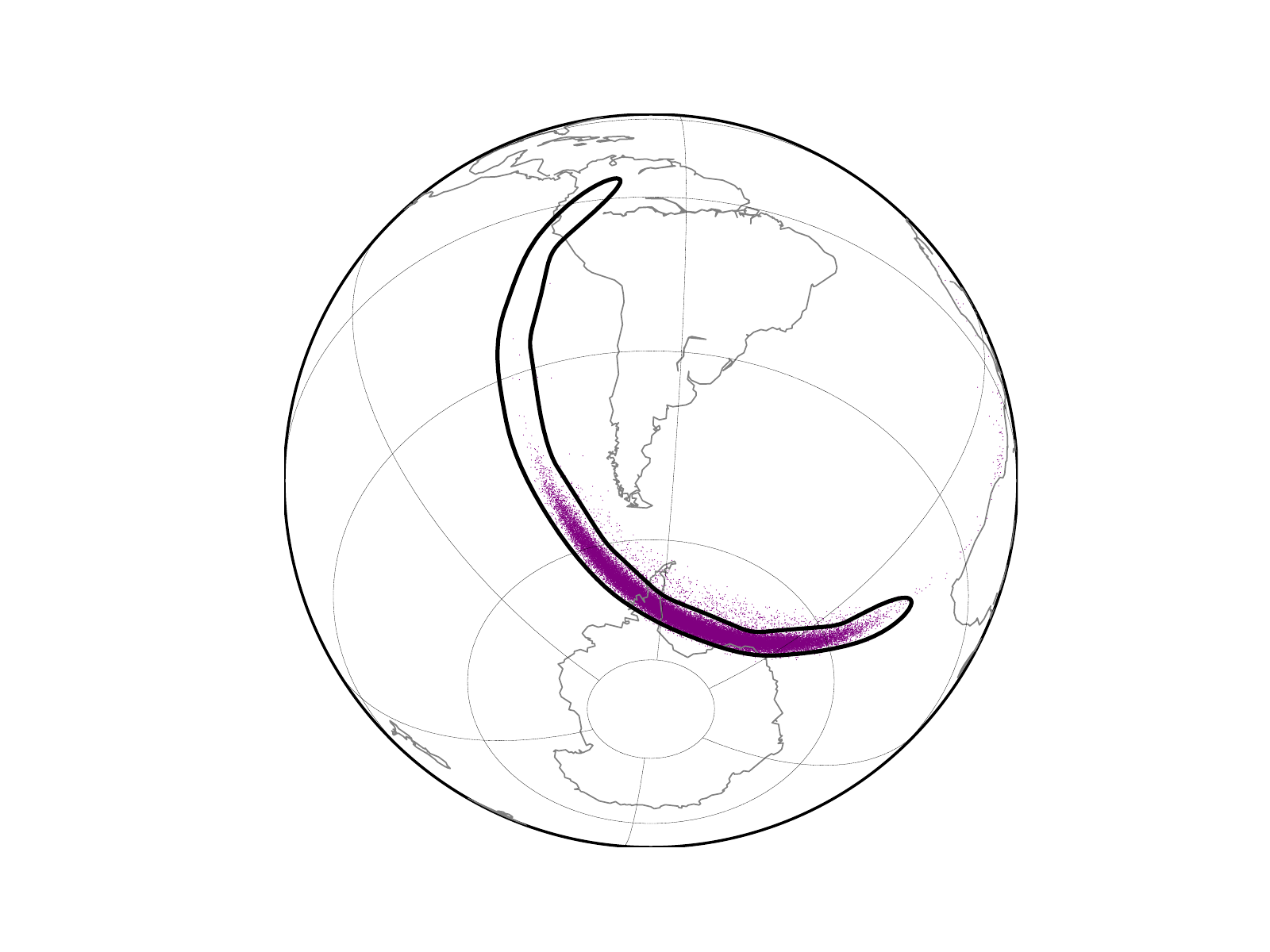}
\caption{\label{fig:sky-post} Orthographic projection of the 90\% two-dimensional contour for the sky position of GW150914 obtained by our
analysis (in black). As a comparison we also show the publicly available posterior samples from a full IMR analysis by the LVC~\cite{O1_samples_release} (in purple, completely overlapping with our result). Regardless of the particular model chosen for the analysis, we always observe the same posterior.}
\end{figure}

\begin{figure}[t]
\center
\includegraphics[width=\columnwidth]{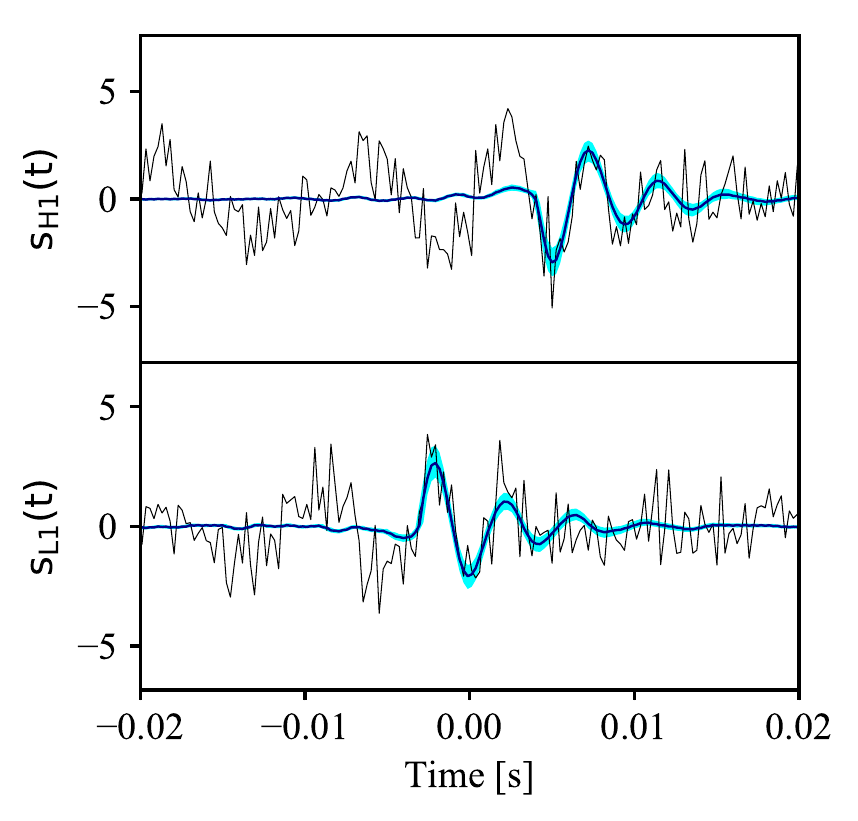}
\caption{\label{fig:reconstructed_wf} Reconstructed whitened waveform superimposed 
on LIGO-Hanford (top panel) and LIGO-Livingston (bottom panel)
data. The solid line shows the median recovered waveform, while 
shaded regions represent $90\% $ credible intervals.}
\end{figure}

The uncertainty on $t_{\rm start}$ is mostly due to the $4096$~Hz sampling
rate used. Interestingly, with a BH mass of $M_f = 68\,M_\odot$ (the median value published by the LVC \cite{TGR_paper_GW150914}), 
one obtains a start time for the ringdown of $\sim 14^{+2}_{-2}\,M_f$.
This result is in good agreement with what obtained through gauge-invariant geometric and 
algebraic conditions quantifying local isometry to the Kerr spacetime~\cite{Swetha} and results obtained 
through earlier parameter estimation methods~\cite{Carullo:2018sfu}.
We also infer a posterior distribution on the sky position of the signal, Fig~\ref{fig:sky-post},
which completely overlaps with published LVC analyses using the full signal~\cite{PE_paper_GW150914, O1_samples_release}.
The projection was obtained using a Dirichlet Process Gaussian-mixture model, as described in \cite{DelPozzo:2018dpu}.
Due to the lower SNR contained in the ringdown-only portion of the signal, our 
posterior distribution is wider. Also, we observe no correlation between $t_{\rm start}$ and 
sky position parameters. 
Fig.~\ref{fig:reconstructed_wf} shows the reconstructed signal overlaid on interferometric data.
A whitening procedure is applied in order to facilitate the visualization of the result, 
but no whitening is applied during the analysis. 

At this point it is natural to ask whether the data provides evidence for a second ringdown mode.
To verify this hypothesis we repeat the previous analysis using the aforementioned settings, 
but now using two independent damped sinusoids.
The Bayes' factor, see Table~\ref{tab:logBs}, shows no evidence for more than a single mode.
We also attempted a test of GR through the measurement of $\delta \omega$ following~\cite{Gossan}, 
but the posterior was uninformative, because the SNR is not enough to constrain more than a single mode. 
Finally, a preliminary study on numerical relativity solutions showed no challenges in testing the no-hair 
conjecture in the high SNR limit, contrary to the claim presented in~\cite{Thrane:2017lqn} and
confirming the results presented in \cite{Brito:2018rfr, Baibhav:2017jhs, Carullo:2018sfu}.

\section{Single Kerr mode} 
From the spectroscopic analysis the favoured modes are the (3,-3,0), (3,-2,0), (2,1,0) and (2,2,0). 
A clear mode identification would require the width of the agnostic posterior to overlap with only a single mode, but
the statistical uncertainty is too large. The GW150914 inspiral result 
points to an almost face-off and nearly equal mass BBH, thus 
the (2,2,0) mode should be the most excited one.
Hence, we wondered whether a stronger assumption 
helps discriminate between the various modes.
We run the analysis with a Kerr model:

\be\label{eq:Kerr_model}
h_+ - i h_{\times} = \frac{M_f}{D_L}\sum_{lmn} \mathcal{A}_{lmn} \, S_{lmn}(\iota, \varphi) \,
e^{i(t-t_{lmn})\tilde{\omega}_{lmn}+\phi_{lmn}}\,,
\ee
where $\tilde{\omega}_{lmn} = \omega_{lmn} + i/\tau_{lmn}$ is the complex ringdown frequency 
determined by the remnant BH mass $M_f$ and its spin $a_f$. The relations  
$\omega_{lmn} = \omega_{lmn}(M_f, a_f)\, ,\tau_{lmn} = \tau_{lmn}(M_f, a_f) $ follow the formulae given in 
Ref.~\cite{LISA_spectroscopy}, available at ~\cite{Berti_website}.
$S_{lmn}$ are the spin-weighted spheroidal harmonics~\cite{Berti:2014fga}.
On the other hand, while analytical predictions for the amplitudes $\mathcal{A}_{lmn}$ exist~\cite{Kamaretsos, MMRDNS_paper, MMRDNP, Baibhav:2017jhs} based on the progenitors masses and spins, 
we do not use them as we wish to estimate them from the data, 
allowing observational comparisons to the aforementioned theoretical models.
Finally, as in the previous analysis, the start time of each mode is left as a free parameter. 
The prior distribution was uniform for parameters:
luminosity distance $D_L\in[10,1000]$\,Mpc, inclination $\cos(\iota)\in[-1,1]$, $M_f\in[10,100]\,\mathrm{M}_{\odot}$, $a_f\in[0, 0.99]$, $\log_{10}{A_{lmn}}\in [-5, 3]$,
$\varphi_i\in[0,2\pi]$\,rad, $t_i\in[3.3,6.6]$\,ms.

In Table~\ref{tab:logBs} we report, for the most probable modes, the Bayes factors comparing the hypotheses that GW150914
can be described as the ringdown generated by a single given $(\ell,m,n)$ Kerr mode.
Without imposing any restriction on the excitations of the different modes, the given 
SNR does not allow us to conclusively discriminate between a subset of the $\ell=2,3$
modes. The obtained posterior on the sky position parameters and the start time show no appreciable difference
with respect to the results already presented in the single damped sinusoid case. 
The posterior distribution of the orientation parameters $(D_L, cos(\iota))$ does not show any 
significant departure from the prior distribution. This is an indication that, with our model, the event is not 
loud enough to infer these parameters from the final stage of the coalescence only.

\section{Kerr multiple modes} 
\begin{figure}[t]
\includegraphics[width=0.45\textwidth]{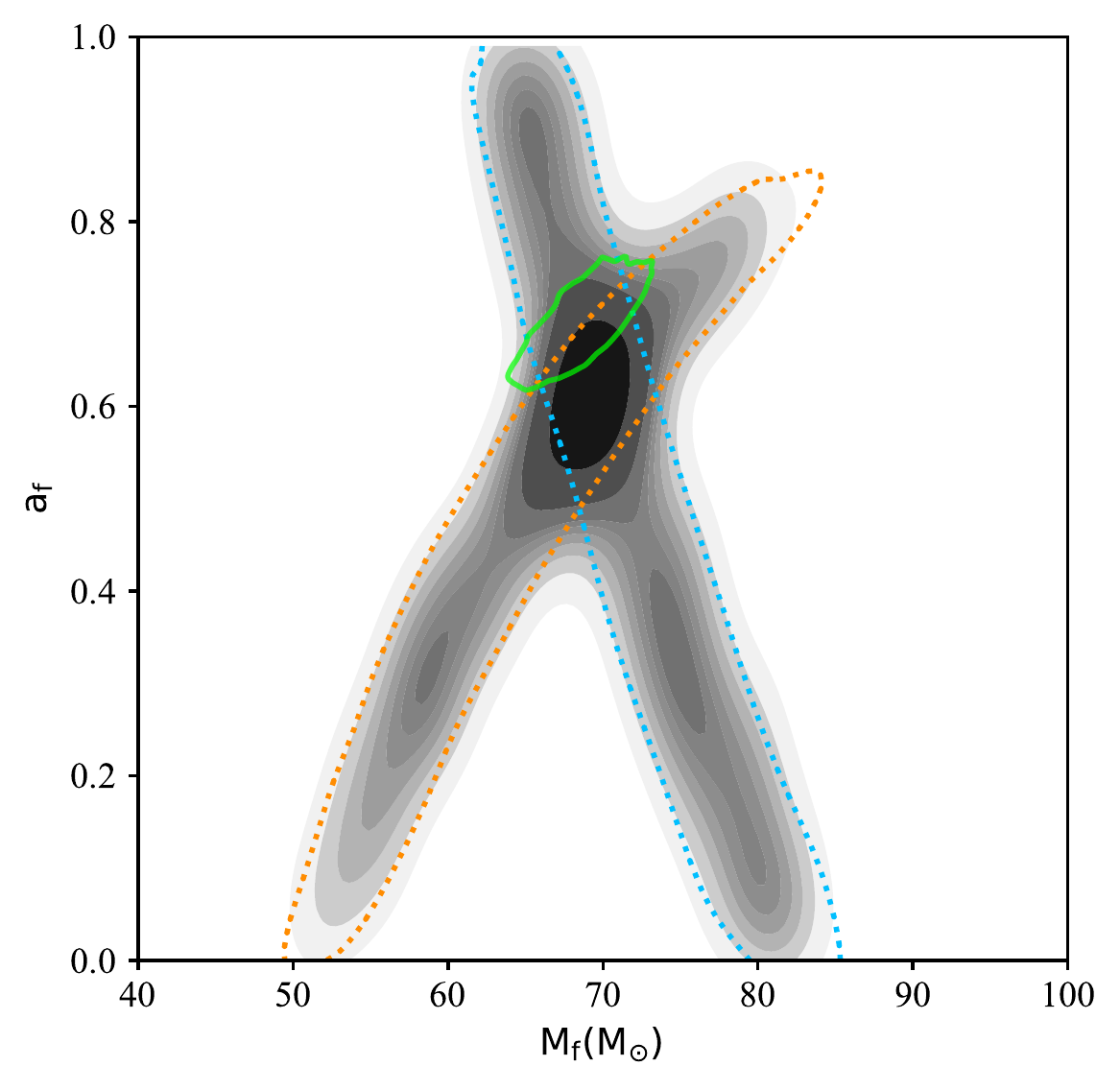}
\caption{\label{fig:mfaf} Posterior distribution for $M_f$ and $a_f$ assuming the presence of individual modes: $(\ell,m,n) = (2,2,0)$ (dotted orange line) \textit{or} $(\ell,m,n) = (3,-3,0)$ (blue dotted line), and from a two-modes analysis $\{ (2,2,0), \, (3,-3,0) \}$ whose posterior density is represented in shades of gray. In green, 
the posterior for $M_f$ and $a_f$ from the LVC IMR analysis. The inferred mass and spin are consistent among the three cases considered, as well as with the LVC posteriors, in line with the Bayes factors, Table~\ref{tab:logBs}, which indicate no preference towards any of them. See the text for a more in depth explanation.}
\end{figure}

Although the unmodelled analysis found no evidence for more than one mode, 
we repeat the analysis with the Kerr model allowing for the presence of two modes to see whether 
a more constraining model is able to detect them.
Table~\ref{tab:logBs} reports results for the few combinations that gave the highest Bayes factors. 
The results on the two Kerr parameters (mass and spin) from this analysis using 
the $\{ (2,2,0), (3,-3,0)\}$ modes is presented in Figure~\ref{fig:mfaf}. 
We wish to stress that the posteriors therein do not imply the presence of multiple modes, but rather explain why the Bayes factors
indicate that we cannot distinguish between a pure $(2,2,0)$, a pure $(3,-3,0)$ or even a mixture of the two, see Table~\ref{tab:logBs}.
Both modes, in fact, provide similar predictions for central frequency and damping time for GW150914 -- see also Fig.~\ref{fig:Spectroscopy_plot} -- for the typical remnant parameters expected from near-equal mass merging BHs, where $a_f\sim 0.6$ is
dominated by the contribution of the orbital angular momentum~\cite{PhysRevD.77.026004, af_paper}. 
From the spectral content only, the inability to discriminate among subsets of modes 
is thus likely to persist in future ground-based observations. We expect this degeneracy to be lifted either in very loud events
or in systems for which the remnant spin is not dominated by the orbital angular momentum. We also note
that, regardless of the final state details, the best systems for spectroscopic studies will be the ones for which 
the orbital configuration is such that the dominant ringdown mode will be the $(2,-2,0)$.

Up to now we kept a semi-phenomenological approach, but a more motivated choice of 
mode combination would be (for example) the use of all $m$ modes for a given $\ell$, 
e.g. for $\ell=2 \,$ $\{ (2,2,0), (2,1,0), (2,0,0), (2,-1,0), (2,-2,0) \}$.
This more generic model implies a much larger number of parameters to be sampled and,
consequently looser bounds.
Results for runs using all the $\ell = 2,3$ modes are summarized in Table~\ref{tab:logBs}
and show no conclusive preference towards one specific $\ell$.
In all cases, the sky position posterior distribution was found to be completely overlapping with the published LVC result, 
while $D_L$ and $\iota$ cannot be well-estimated due to the complete degeneracy with mode amplitudes.
The posterior on start time shows minimal changes with respect to 
the one obtained by the damped-sinusoid analysis, for all the employed models.
Its stability  is a strong indication of the robustness of our method, since no variation between different modes start times
are expected to be detectable with the SNR contained in the post-merger portion of GW150914.

\begin{table}
\caption{Summary of the Bayes factors, $M_f, a_f$ median (when measured) and 90$\%$ CI obtained with different waveform models (DS stands for Damped Sinusoid) and
a transition time prior in $[3.3,6.6]$\,ms. The statistical errors on the log Bayes' factors are $\pm \, 0.1$. They are estimated computing their variance 
over 10 different realisations of the pseudo-random number 
chain initialisation. Within the statistical errors, when differences of $\log B$ are compared against heuristic evidence scales, 
such as the Jeffreys scale, no significant evidence in favor of any specific mode (or combination of modes) is present.}
\begin{ruledtabular}
\begin{tabular}{llll}
Model & logB$_{\mathrm{s,n}}$ & $M_f/M_\odot$ & $a_f$\\
\hline
IMR (LVC) 						 & -  		  		  & $68.0^{+\, 3.2}_{-\, 3.0}$   & $0.69^{\, 0.05}_{\, 0.04}$  \\
DS - 1 					   mode  & $56.3$ &  - 			  		       &  -   					   \\
DS - 2       			   modes & $55.4$ &  -   			  		   &  -  		       		   \\
Kerr - (2,2,0)		   	   mode  & $56.5 $ & $64.6^{+\, 14.3}_{-\, 11.4}$ & $0.50^{+\, 0.28}_{-\, 0.40}$ \\
Kerr - (2,1,0)			   mode  & $56.6$ & $61.2^{+\, 8.9}_{-\, 8.5}$   & $0.60^{+\, 0.28}_{-\, 0.49}$ \\
Kerr - (2,0,0)			   mode  & $56.0$ & $55.0^{+\, 4.1}_{-\, 4.1}$   & $0.69^{+\, 0.27}_{-\, 0.58}$ \\
Kerr - (3,-3,0) 			   mode  & $57.2$ & $72.3^{+\, 9.7}_{-\, 8.1}$   & $0.46^{+\, 0.47}_{-\, 0.42}$ \\
Kerr - (3,-2,0)			   mode  & $57.0$ & $75.7^{+\, 7.1}_{-\, 5.5}$   & $0.49^{+\, 0.44}_{-\, 0.43}$ \\
Kerr - (3,-1,0) 			   mode  & $57.0$ & $79.9^{+\, 4.5}_{-\, 3.8}$   & $0.47^{+\, 0.46}_{-\, 0.43}$ \\
Kerr - (2,2,0),(3,-3,0)    modes  & $56.7$ & $69.2^{+\, 12.1}_{-\, 14.2}$ & $0.50^{+\, 0.40}_{-\, 0.44}$ 	\\
Kerr - (2,2,0),(2,1,0)     modes  & $56.2$ & $62.7^{+\, 15.6}_{-\, 9.9}$  & $0.54^{+\, 0.31}_{-\, 0.44}$  \\
Kerr - $\ell=2$ 			   modes & $55.0$ & $55.1^{+\, 15.5}_{-\, 7.9}$  & $0.53^{+\, 0.54}_{-\, 0.46}$  \\
Kerr - $\ell=3$ 			   modes & $54.3$ & $81.9^{+\, 13.2}_{-\, 10.5}$  & $0.31^{+\, 0.54}_{-\, 0.28}$  \\
Kerr - $\ell=2,3$ 		   modes & $52.0$ & $56.6^{+\, 27.9}_{-\, 10.1}$  & $0.39^{+\, 0.47}_{-\, 0.36}$  \\
\end{tabular}
\label{tab:logBs}
\end{ruledtabular}
\end{table}

\begin{figure*}[t]
\includegraphics[width=1\textwidth]{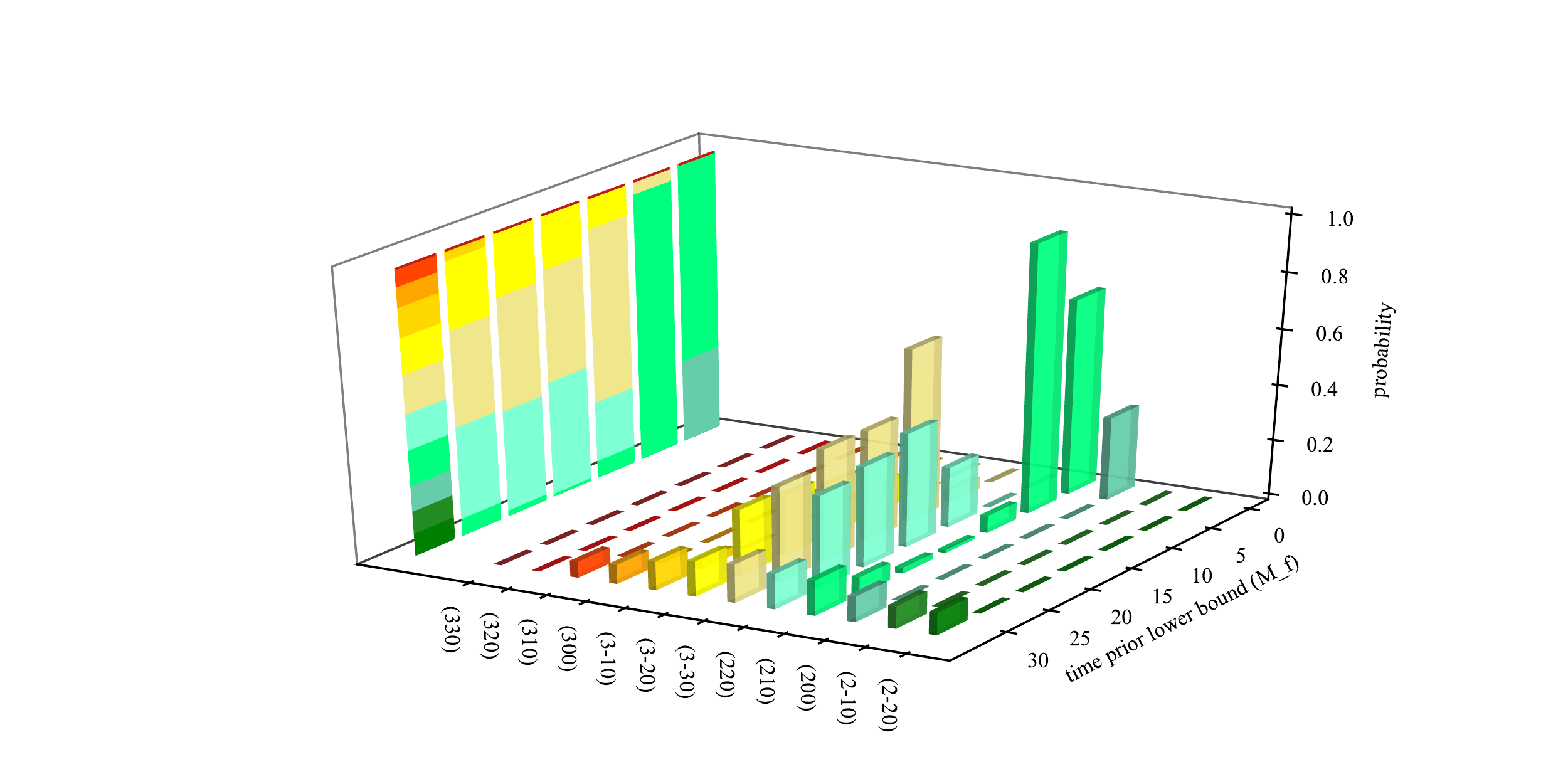}
\caption{\label{fig:3dprobs} Mode identification probability as a function of the lower prior bound on the transition time. 
Each histogram represents the probability that the identified frequency and 
damping time correspond to the $k$-th mode as predicted assuming the LVC final mass and spin measurements.
The color scheme is the same as in Fig.~\ref{fig:Spectroscopy_plot} to ease comparison.
As expected, starting the analysis at early times returns frequencies that are lower than that expected for the
(2,2,0) mode, and lead to other modes being preferred.
In particular, the (2,1,0) mode is the most probable until the start time is smaller than $10\,M_f$ (green histogram). 
For times between $10\,M_f$ and $25\,M_f$ the most probable modes are the $(2,2,0)$, the $(3,-3,0)$ and the $(3,-2,0)$, 
aquamarine, khaki and yellow histograms, respectively.
Finally, for times greater than $25\,M_f$ the GW150914 signal is too quiet to reliably identify any specific mode.
The left most bars summarise the mode probabilities.}
\end{figure*}

\begin{figure}[h]
\includegraphics[width=0.45\textwidth]{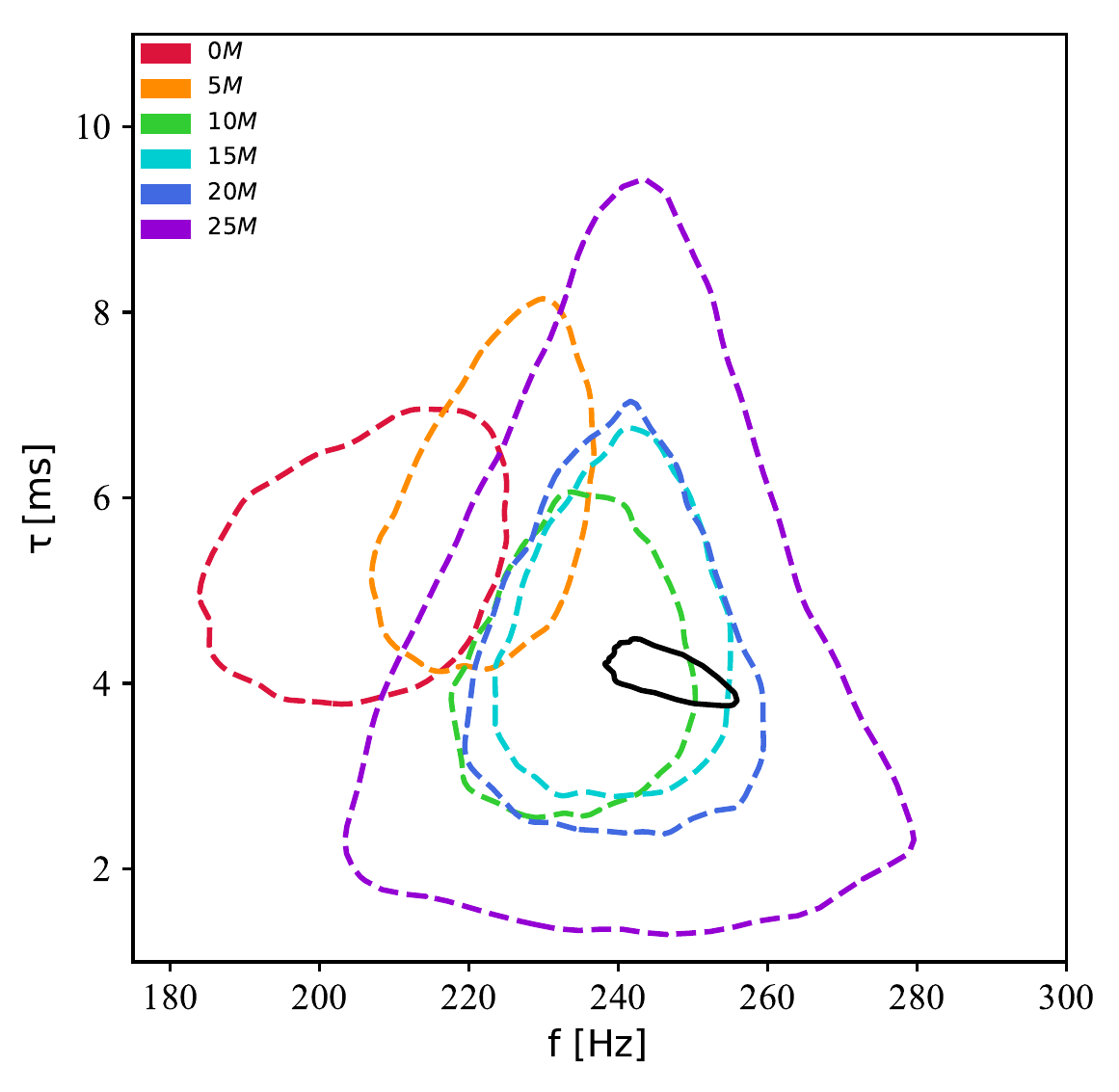}
\caption{\label{fig:freqs_time_prior} Effect of the start time prior on the posterior onfthe reconstructed frequency and damping time, when assuming a single damped sinusoid. The solid black band marks the prediction obtained by using the LVC posteriors and assuming the excited mode was the $(\ell,m,n) = (2,2,0)$. Using a lower bound on the start time of $30 \, M_f$ the obtained reconstruction is flat, thus it is not shown in the plot.}
\end{figure}

\begin{figure}[h]
\includegraphics[width=0.45\textwidth]{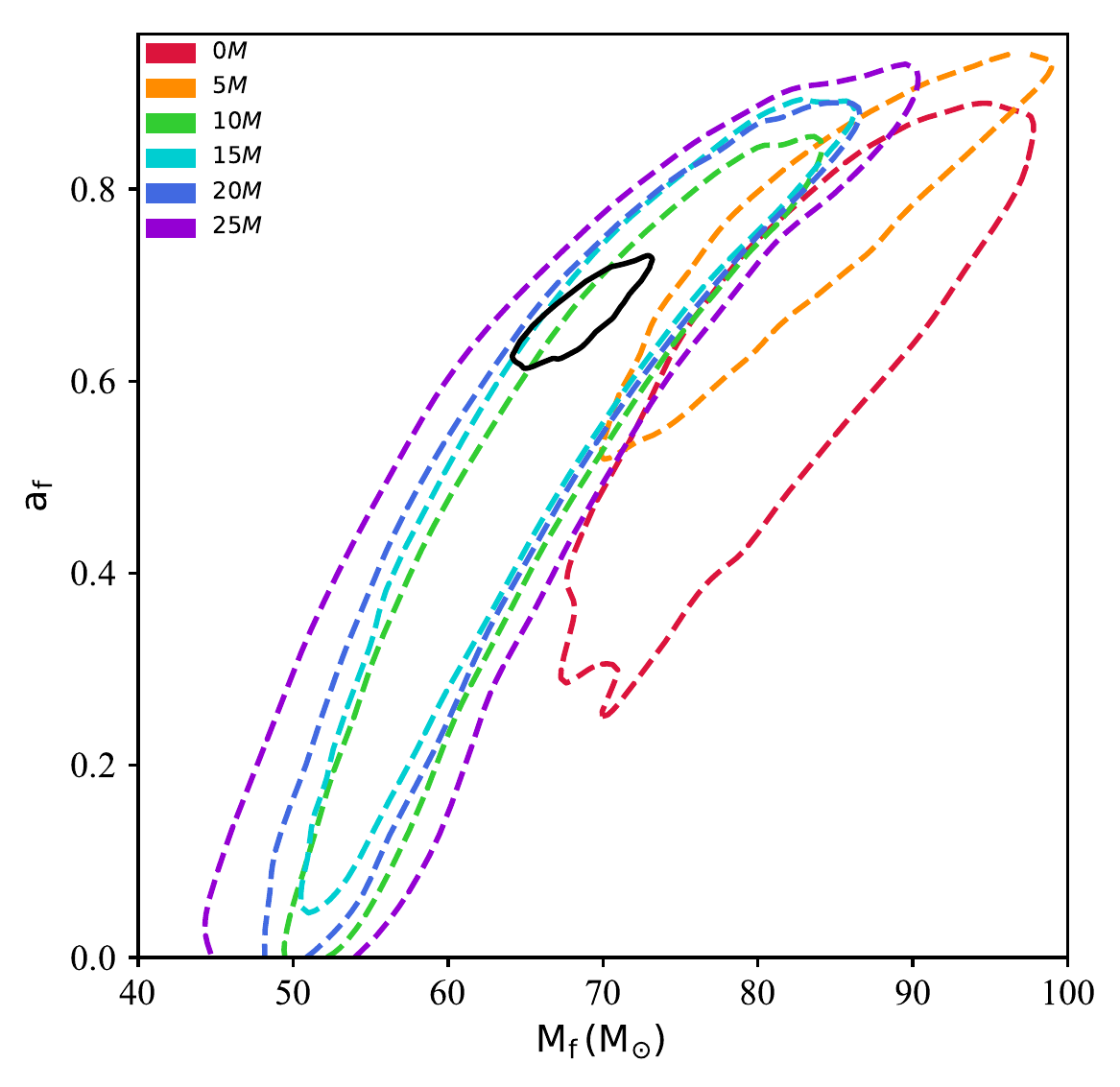}
\caption{\label{fig:mass_time_prior} Effect of the start time prior on the posterior onfthe reconstructed final mass and spin, when assuming a $(\ell,m,n) = (2,2,0)$ Kerr mode. The solid black band marks the prediction obtained by using the LVC posteriors. Using a lower bound on the start time of $30 \, M_f$ the obtained reconstruction is flat, thus it is not shown in the plot.}
\end{figure}

\section{Discussion}

We now focus on a discussion of some of the key assumptions in our analysis. 
We begin by discussing the effect of varying the lower bound of the prior distribution for the transition time,
which earlier we have taken to be uniform in the range $[10,20]~\mathrm{M}_f$ after the peaktime of $h_+^2 + h_\times^2$. This choice was guided by numerical relativity studies which looked at the validity
of the linearized regime \cite{MMRDNS_paper, Swetha, Carullo:2018sfu},
but it is interesting to explore how our inference is affected by relaxing this assumption.
We do so by varying the lower bounds of the prior from $0\,M_f$ -- corresponding to the peak of the waveform -- to $30\,M_f$ in steps of $5\,M_f$, with prior widths fixed to $10\,M_f$.

We find that the posterior distribution over $t_{\rm start}$ is dependent on the choice of the prior. 
For the earliest time prior considered ($t_{\rm start}\in[0.0,3.3]$\,ms, corresponding to $[0,10]\,M_f$), 
we obtain a measurement of $t_{\rm start}\simeq 1\,$ms. 
For the latest time prior ($t_{\rm start}\in[30,40] \,M_f$), we obtain a flat posterior distribution as the SNR is too small to obtain a measurement.
In all other cases, except the more theoretically guided choice ($t_{\rm start}\in[10,20]\,M_f$),
we find that the posterior rails against the lower bound of the prior; as noted in~\cite{Carullo:2018sfu} 
the template attempts to be as faithful as possible to the data, while maximising 
the recovered signal-to-noise ratio. 
The sky position recovery is consistent with the result obtained by the LVC ~\cite{PE_paper_GW150914, O1_samples_release}
when the lower bound of the prior is strictly smaller than $15M_f$. Starting from $15M_f$ onwards,
the obtained sky position is biased, reflecting the fact that, at this low SNR, the template tries to latch as early 
as possible to the data (within the given time prior bounds).
The results discussed above are independent of the specific waveform model employed.

The sensitivity to the time prior choice is reflected also in the recovery of the intrinsic parameters of the binary.
When using a superposition of damped sinusoids, it will affect mode identification:
Fig.~\ref{fig:3dprobs} shows how the mode identification varies as a function of the time prior. As the lower bound of the time prior approaches the waveform peak,
the recovered frequency becomes smaller and smaller due to the waveform latching onto the merger. As consequence, initially, the most probable mode is 
the $(2,1,0)$ (green histogram). For intermediate times, the situation is the one presented in Fig.~\ref{fig:Spectroscopy_plot} where the most probable modes are the 
$(2,2,0)$, the $(3,-3,0)$ and the $(3,-2,0)$, aquamarine, khaki and yellow histograms, respectively. For start times greater than  $25\,M_f$ no inference is possible anymore and all modes become essentially equally probable. 
In Fig. \ref{fig:freqs_time_prior} we show the dependence of the posterior distribution on the frequency and damping time on the time prior.

When considering a Kerr template\footnote{Here we choose the $(2,2,0)$ mode as a representative case, 
but the considerations presented are common to all the used modes}, 
Fig.~\ref{fig:mass_time_prior} shows how the reconstruction of mass and spin varies as a function of the time prior lower bound.
As already discussed, earlier times correspond to smaller recovered frequencies and consequently to larger masses.
In all the considered cases the orientation parameter ($D_L$, $\iota$) posterior distribution shows no departure from its prior distribution, due to the degeneracy with the mode excitation amplitude $\mathcal{A}_n$.

In conclusion, if GR indications on where a ringdown model with a dominant (2,2,0) mode is valid are taken into account to set up a prior on the start time, the recovered parameters of the binary are consistent with the predictions of GR.
Moreover a measurement of the effective start time of the ringdown regime can be achieved, excluding both early and late times within the given prior bounds.
If later times are allowed by the prior, then the corresponding time posterior rails against its lower bound,
signaling a preference towards including the earlier portion of the signal in the analysis.
If instead the analysis is performed starting from the peak of the waveform, the recovered parameters are biased with respect to the GR prediction, which we interpret as signaling that the system was in a regime where the considered ringdown models do not provide a good description of the emitted radiation. 
Recent work with NR simulations has suggested that the inclusion of overtones in the model (i.e. $n\neq0$ modes) allow the ringdown to be extended back toward $t_\mathrm{start}$ without biasing other parameters~\cite{Giesler:2019uxc}. This will be the subject of a subsequent investigation.

\section{Summary} 
We presented the first comprehensive spectroscopic study of GW150914 ringdown.
For the first time, we measured the ringdown onset time from the data, 
and we found it to be consistent with predictions from numerical relativity.
Bayesian model selection indicates that the data do not provide evidence in support of the presence of multiple QNM. We attempted to identify 
which QNM is dominant, obtaining the (3,-3,0), (3,-2,0), (2,1,0) and (2,2,0) modes as the most probable ones.
At the GW150914 SNR, we cannot determine univocally the QNM label, in agreement with 
preliminary results on numerical relativity waveforms not presented here for brevity. 
More targeted investigations based on the Kerr BH solution, confirm our model-independent findings as well as
our current lack of mode-resolving power. We note that, due to the similar mode
frequencies for the aforementioned subset of modes excited in a near-equal-mass BH coalescences with small spins, 
BH spectroscopy with a moderate SNR will require the use of information 
from the inspiral phase to determine the most probable modes. 
A systematic investigation of the details of our method applied to numerical waveforms, 
together with the inclusion of refined waveform models 
incorporating GR numerical predictions, will be presented in a future study.
The analysis presented in this paper can and will be 
applied to louder and/or multiple GW events. Joint coherent analyses will help to test the predictions of linearized GR. 
Such an extension of the present work will be presented in a future publication.
We also defer to a further publication a detailed analysis measuring relative deviations from GR ringdown frequencies~\cite{Gossan, Meidam}.

\begin{acknowledgments}
\textit{Acknowledgments}
The authors would like to thank Lionel London for providing a fit of the spheroidal harmonics. We are grateful to Abhirup Ghosh for useful discussions and to Vivien Raymond for suggestions on the manuscript. This work greatly benefited from discussions within the \textit{strong-field} working group
of the LIGO/Virgo collaboration.
W.D.P. is funded by the ``Rientro dei Cervelli Rita Levi Montalcini'' Grant of the Italian MIUR.
JV was supported by STFC grant ST/K005014/2. This work made use of the ARCCA Raven cluster, funded by STFC grant ST/I006285/1 supporting UK Involvement in the Operation of Advanced LIGO.
 This research has made use of data, software and/or web tools obtained from the Gravitational Wave Open Science Center (https://www.gw-openscience.org), a service of LIGO Laboratory, the LIGO Scientific Collaboration and the Virgo Collaboration. LIGO is funded by the U.S. National Science Foundation. Virgo is funded by the French Centre National de Recherche Scientifique (CNRS), the Italian Istituto Nazionale della Fisica Nucleare (INFN) and the Dutch Nikhef, with contributions by Polish and Hungarian institutes.
\end{acknowledgments}

\bibliographystyle{apsrev}
\bibliography{RingdownTD_Bibliography}

\end{document}